\documentclass[prd,twocolumn,floatfix,preprintnumbers,letterpaper]{revtex4}
\usepackage{graphicx}
\usepackage{amsmath,amssymb}
\input{epsf}
\usepackage{epsf}

\begin{document}

\def\be{\begin{equation}}
\def\ee{\end{equation}}

\title{Phantom Dark Energy Models with Negative Kinetic Term}
\author{Jens Kujat,$^{1,2}$ Robert J. Scherrer,$^{1}$ and A.A. Sen$^1$}
\affiliation{$^1$Department of Physics \& Astronomy, Vanderbilt University,
Nashville, TN~~37235}
\affiliation{$^2$Department of Physics, The Ohio State University,
Columbus, OH~~43210}
\date{\today}

\begin{abstract}
We examine phantom dark energy models derived from a scalar field
with a negative kinetic term for which
$V(\phi) \rightarrow
\infty$ asymptotically.
All such models can be divided into three classes, corresponding
to an equation of state parameter $w_\phi$ with asymptotic behavior
$w_\phi \rightarrow -1$, $w_\phi \rightarrow w_0 < -1$, and $w_\phi \rightarrow
- \infty$.  We derive the conditions on the potential $V(\phi)$ which
lead to each of these three types of behavior.  For models with
$w_\phi \rightarrow -1$, we derive the conditions on $V(\phi)$ which determine
whether or not such models produce a future big rip.  Observational
constraints are derived on two classes of these models:  power-law
potentials with $V(\phi) = \lambda \phi^\alpha$
(with $\alpha$ positive or negative) and exponential potentials
of the form $V(\phi) = \beta e^{\lambda \phi^\alpha}$. 
It is shown
that these models spend more time in a
state with $\Omega_m \sim \Omega_\phi$ than do corresponding models with
a constant value of $w_\phi$, thus providing a more satisfactory
solution to the coincidence problem.
\end{abstract}

\maketitle

\section{Introduction}

Observational evidence \cite{Knop,Riess}
indicates that roughly 70\% of the energy density in the
universe is in the form of an exotic, negative-pressure component,
dubbed dark energy.  (See Ref. \cite{Copeland}
for a recent review).  If $\rho_\phi$ and $p_\phi$ are the density
and pressure, respectively, of the dark energy, then
the dark energy can be characterized by the
equation of state parameter $w_\phi$, defined by
\begin{equation}
w_\phi = p_\phi/\rho_\phi.
\end{equation}

It was first noted by Caldwell \cite{Caldwell} that
observational data do not rule out the possibility
that $w_\phi <-1$.  Such ``phantom" dark energy
models have several peculiar properties.
The density of the dark energy {\it increases} with
increasing scale factor, and both the scale factor
and the phantom energy density can become infinite
at a finite $t$, a condition known as the ``big rip" \cite{Caldwell,rip,rip2}.
Further, it has been suggested that the finite lifetime for the universe
which is exhibited in these models may provide an explanation for
the apparent coincidence between the current values of the
matter density and the dark energy density \cite{doomsday}.

The simplest way to achieve a phantom model is to take a scalar
field Lagrangian with a negative kinetic term.  While such
models have certain well-known problems \cite{Carroll,Cline,Hsu1,Hsu2},
they
nonetheless provide an interesting set of representative phantom
models, and they have been widely studied \cite{Guo,ENO,NO,Hao,Aref,Peri,Sami,
Faraoni}.

In this paper, we provide a more general
analysis of phantom models with negative kinetic
terms.  In the next section, we reexamine the
behavior of such models, and we generalize
previous studies to determine the asymptotic
behavior of such models and the generic conditions
for the existence of a future singularity.
In Sec. 3, we derive observational
constraints on a subset of these models from the supernova data.
In Sec. 4, we examine the solution to the coincidence problem
for these models.
Our conclusions are summarized in Section 5.

\section{Phantom Models}

\subsection{General Results}

We limit our discussion to a spatially-flat universe, for which
\be
\label{Fried}
H^2 =  \frac{\rho}{3} 
\ee
and
\be
\label{Fried2}
\dot{\rho} =  -3H(\rho + p) \,,
\ee
where $H = \dot a/a$, $a$ is the scale factor,
and we take $\hbar = c = 8\pi G = 1$ throughout.
In equations (\ref{Fried}) and (\ref{Fried2}),
$\rho$ and $p$ represent the total
energy density and pressure of the
matter, radiation, and phantom fields:
\begin{eqnarray}
\rho &=& \rho_m + \rho_r + \rho_{\phi},\\
p &=& p_r + p_{\phi}.
\end{eqnarray}

In a phantom model with negative kinetic
term, the energy density
and pressure of the phantom are given by
\be
\rho_{\phi}={-(1/2)\dot \phi^2 + V(\phi)},
\ee
and
\be
p_{\phi}={-(1/2)\dot \phi^2 - V(\phi)},
\ee
so that the equation of state parameter is
\begin{equation}
\label{w}
w_{\phi} =\frac{(1/2)\dot \phi^2 + V(\phi)}{ (1/2)\dot \phi^2 - V(\phi)}\,,
\end{equation}

The evolution equation for this field is
\begin{equation}
\label{phiev}
\ddot{\phi}  +   3 H \dot{\phi} - V^\prime(\phi) = 0.
\end{equation}
where the prime denotes the derivative with respect
to $\phi$.
A field evolving according to equation (\ref{phiev}) rolls
uphill in the potential.  If the field gets stuck
in a local maximum, we simply have $w_\phi = -1$ and no
future singularity \cite{Carroll,Hao,Faraoni}.  We
will confine our attention to models
for which $V(\phi) \rightarrow \infty$ asymptotically.
(A third possibility
is a phantom model for which $V(\phi)$ approaches a constant,
e.g., a potential of the form $V(\phi) = V_0 \phi/(\phi + \phi_0)$.
We will not examine such models here, but they are discussed briefly
in Ref. \cite{ENO}).

Specific phantom models with $V(\phi) \rightarrow \infty$ have been previously
examined in Refs. \cite{Guo,ENO,NO,Hao,Aref,Peri},
while some conditions
which produce a future singularity have been examined by
Sami and Toporensky \cite{Sami}
and by Faraoni \cite{Faraoni}.  It is these latter
studies which we will now generalize.

Sami and Toporensky considered
several classes of models, including power-law potentials,
$V(\phi) \propto \phi^\alpha$ with $\alpha > 0$, exponential
potentials with $V(\phi) \propto e^{\lambda \phi}$, and potentials
steeper than exponential (specifically, $V(\phi) \propto e^{\lambda \phi^2}$).
They found that the power-law potentials asymptotically produce
$w_\phi \rightarrow -1$, with a big rip occurring only for the cases
with $\alpha > 4$.  Exponential potentials lead to a big
rip with $w_\phi$ approaching a constant (see also
Refs. \cite{ENO,Hao}), while the $e^{\lambda \phi^2}$
potential gives a singularity with $w_\phi \rightarrow - \infty$.

Consider first the general relation between the asymptotic behavior
of $V(\phi)$ and the asymptotic behavior of $w_\phi$.
This relation can be derived by
writing the equation of motion as \cite{Chiba}
\begin{equation}
\label{Vprime/V}
\pm \frac{V^\prime}{V} = \sqrt{\frac{-3(1+w_\phi)}{\Omega_\phi}}\left[1 + \frac{1}{6}
\frac{d\ln(x)}{d\ln(a)}\right],
\end{equation}
where
$\Omega_\phi$ is the density of the
phantom field in units of the critical
density (note that $\Omega_\phi$ evolves with time).
In equation (\ref{Vprime/V}), $x = \dot\phi^2/2V$, so that
$x$
and $w_\phi$ are related via
\begin{equation}
x = - \frac{1+w_\phi}{1-w_\phi}.
\end{equation}
Equation (\ref{Vprime/V}) differs
slightly from the corresponding equation
in Ref. \cite{Chiba} because we use a different
definition of $x$.  Equation (\ref{Vprime/V}) is the phantom
version of the quintessence equation of motion first derived
in Ref. \cite{Steinhardt}; it differs from the latter equation
only in the sign of $1+w$ on the right-hand side.

Now consider the asymptotic evolution of $w_\phi$ in the limit where
the universe is phantom-dominated.  We can distinguish three possible
cases, which are determined by the asymptotic behavior of $V^\prime/V$
as $V \rightarrow \infty$.  Note first that if $V^\prime/V$ asymptotically
approaches a finite constant, a solution exists for which
$w$ approaches a constant; in this case we have $d\ln(x)/d\ln(a) \rightarrow
0$ in equation (\ref{Vprime/V}).  On the other hand,
if $V^\prime/V$ becomes asymptotically infinite, we see
that there is a solution with $w \rightarrow - \infty$.  In
this case, $x \rightarrow 1$ (the latter result was noted in
the numerical simulations of Ref. \cite{Sami}), so that
we again have
$d\ln(x)/d\ln(a) \rightarrow
0$ in equation (\ref{Vprime/V}).  Thus, from
equation (\ref{Vprime/V}), the relation between
$V^\prime/V$ and $w_\phi$ is given by:

\vskip 0.4 cm

\noindent Class 1:
\be
\frac{V^\prime}{V} \rightarrow 0  \Leftrightarrow w_\phi \rightarrow -1,
\ee

\noindent Class 2:
\be
\frac{V^\prime}{V} \rightarrow constant  \Leftrightarrow w_\phi \rightarrow w_0< -1,
\ee

\noindent Class 3:
\be
\frac{V^\prime}{V} \rightarrow \pm\infty  \Leftrightarrow w_\phi \rightarrow -\infty.
\ee

These results illuminate the behavior of the specific examples in
Refs. \cite{Hao,Sami}.  For an arbitrary power-law potential,
$V \propto \phi^\alpha$, we have
$V^\prime/V  = \alpha/\phi$.  For positive $\alpha$,
$\phi \rightarrow \infty$ and $V^\prime/V \rightarrow 0$,
so we have $w_\phi \rightarrow -1$, as noted in Ref. \cite{Sami}.
On the other hand, if $\alpha < 0$, as in the models investigated below,
we have $\phi \rightarrow 0$ and $V^\prime/V \rightarrow -\infty$,
so that $w_\phi \rightarrow -\infty$.

The only potential that corresponds exactly to Class 2 is the
exponential:  $V(\phi) \propto e^{\lambda \phi}$.  In this case
we see that $w_\phi$ approaches a constant; equation (\ref{Vprime/V}) gives
$w_\phi = -1 - \lambda^2/3$, in agreement with the results of Ref. \cite{Hao}.

Finally, the potential $V \propto e^{\lambda \phi^\alpha}$ gives
$V^\prime/V = \lambda \alpha \phi^{\alpha-1}$.
For $\lambda > 0$ and $\alpha > 0$, the field rolls in the positive
$\phi$ direction, and we see that $V^\prime/V \rightarrow 0$
for $\alpha < 1$ (so $w_\phi \rightarrow -1$), while $V^\prime/V \rightarrow
\infty$ for $\alpha > 1$, so $w_\phi \rightarrow -\infty$.
For the particular case $\alpha = 2$, Sami and Toporensky
\cite{Sami} found $w_\phi \rightarrow - \infty$, which agrees with our
general result.

Faraoni \cite{Faraoni} showed that
models in which $V(\phi)$ has a vertical asymptote necessarily lead
to a singularity in a finite time.  All such models fall under our Class 3
(e.g., negative power-law potentials), but there are also models
in Class 3 that do not have a vertical asymptote
(e.g., $V(\phi) \propto e^{\lambda \phi^\alpha}$ with $\lambda > 0$ and
$\alpha > 1$).

All models belonging to Class 2 and Class 3 necessarily lead to a
future singularity,
but
for Class 1, we must distinguish models that produce a big
rip from those that do not.
In the limit where $w_\phi \rightarrow -1$,
we have $\dot \phi^2/2V \rightarrow 0$, and we can neglect
the $\ddot\phi$ term in equation (\ref{phiev}), so that equations
(\ref{phiev}) and (\ref{Fried}) reduce to \cite{Sami}
\begin{equation}
\dot \phi = \frac{V^\prime(\phi)}{3H},
\end{equation}
and
\begin{equation}
H^2 = \frac{V(\phi)}{3}.
\end{equation}
Combining these two equations gives
\be
\label{dotphi}
\dot \phi = \frac{V^\prime(\phi)}{\sqrt{3V(\phi)}}.
\ee
The condition for a big rip is that $V(\phi)$
become infinite in a finite time.  From equation (\ref{dotphi}),
this big rip condition is equivalent
to the requirement that
\be
\label{ripcondition}
\int \frac{\sqrt{V(\phi)}}{V^\prime(\phi)} d\phi \rightarrow finite
\ee
One limit of integration in equation (\ref{ripcondition})
is the value of $\phi$ for which $V \rightarrow \infty$,
while the other limit is fixed at an arbitrary constant
$\phi_c$.  Equation (\ref{ripcondition}) distinguishes models
with $w_\phi \rightarrow -1$ that produce a big rip from those that
do not.

Now consider the specific example of the positive power-law
potentials.  For $V(\phi) = \lambda \phi^\alpha$, with $\alpha > 0$,
the integral in equation (\ref{ripcondition}) becomes
\be
\int_{\phi_c}^\infty
\frac{\phi^{\alpha/2}}{\sqrt{\lambda}\alpha\phi^{\alpha-1}}d\phi
\propto\phi^{2-\alpha/2}|_{\phi_c}^\infty
\ee
for $\alpha \ne 4$.
Clearly, this integral converges (corresponding to a big rip)
for $\alpha > 4$ and diverges (giving no big rip) for $\alpha < 4$,
in agreement with the results of Ref. \cite{Sami}.  The case $\alpha =4$
gives an integral that diverges logarithmically, corresponding
to no big rip.  Again, this agrees with Ref. \cite{Sami}.

The future singularity that occurs for Class 1 and Class 2
has the property that the scale factor and density both become infinite
at a finite time, $t_s$.  (In
the classification scheme
of Nojiri, Odintsov, and Tsujikawa \cite{NOT}, this is a Type I singularity).
This need not be the case for Class 3, in which $w_\phi \rightarrow -\infty$.
As noted in Ref. \cite{Sami}, the potential
$V \propto e^{\lambda \phi^2}$ produces a singularity
in which the scale factor goes to a constant at finite $t_s$, but
the density becomes infinite (a Type III singularity in Ref. \cite{NOT}).
We find that the negative power models investigated
in the next section also produce a singularity with finite
scale factor and infinite density.
\begin{figure}[t]
\centerline{\epsfxsize=3truein\epsffile{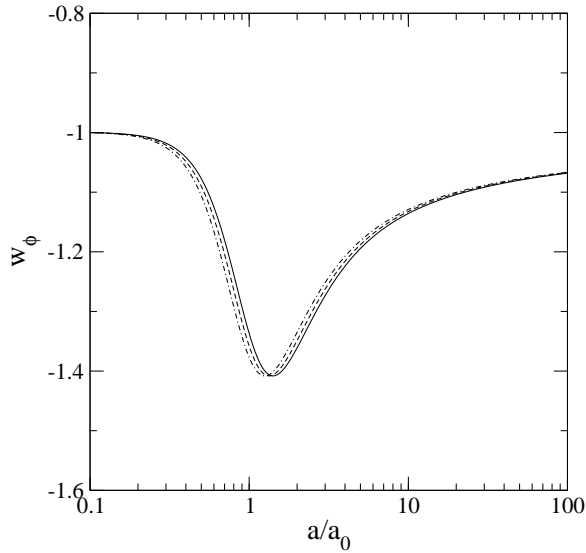}}
\caption{Evolution of the equation of state parameter $w_{\phi}$ as a function of
scale factor $a$ ($a_0$ is the scale factor today)
for $V = \lambda \phi^2$ with $\Omega_{m0} = 0.25$ (dot-dashed curve),
$\Omega_{m0} = 0.3$ (dashed curve) and $\Omega_{m0} = 0.35$ (solid curve),
with $\phi_0 = 1$.}
\end{figure}

\subsection{Specific Cases}
 
\subsubsection{Power-Law Potentials}
 
Consider first the case of power-law potentials,
for which
\begin{equation}
\label{power}
V(\phi) = \lambda \phi^{\alpha},
\end{equation}
and we allow for the exponent $\alpha$ to be either positive or negative.
Positive power-law potentials have been previously investigated in Refs.
\cite{Sami,Peri,Guo}.  Negative power-law potentials have not been previously
investigated in general, although Hao and Li \cite{Hao} showed
that such models do not lead to tracker solutions.
Note that $\lambda$ in equation (\ref{power}) is chosen to fix the value of $\Omega_\phi$ today, so it is
not taken to be a free parameter in these models.

We evolve these models using equation (\ref{phiev}).
\begin{figure}[t]
\centerline{\epsfxsize=3truein\epsffile{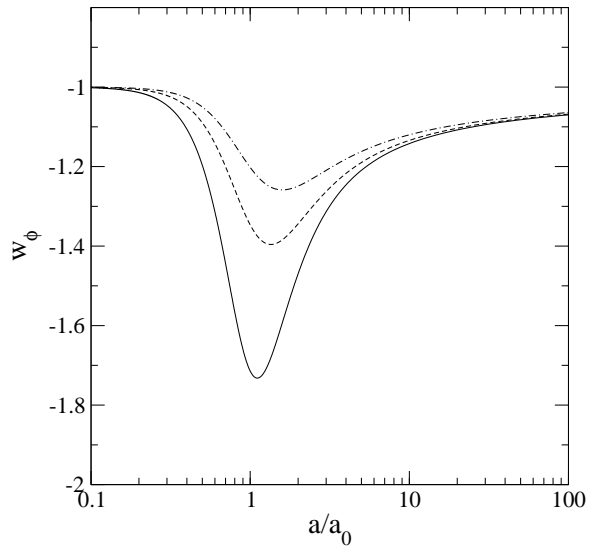}}
\caption{Evolution of the equation of state parameter $w_{\phi}$ as a function of
scale factor $a$ ($a_0$ is the scale factor today)
for $V = \lambda \phi^2$ with
$\Omega_{m0} = 0.3$, and
$\phi_0 = 0.5$ (solid curve), $\phi_0 = 1$ (dashed curve), and $\phi_0 = 1.5$
(dot-dashed curve).}
\end{figure}
We
take $\dot \phi = 0$ initially, but we have verified that
the evolution is independent of this initial condition.
As long as $\rho_\phi \ll \rho_m + \rho_r$,
$\dot \phi$ quickly
decays to zero.
On the other hand, we do find that these models are highly
sensitive to the initial value for $\phi$, which we designate
as $\phi_0$.  This contrasts sharply with the behavior of
ordinary quintessence models with power-law potentials, for which
the evolution can be insensitive to the initial conditions \cite{Steinhardt,LS}.
The actual value of $\phi_0$ is, of course, a model-dependent quantity.
(For quintessence models,
some estimates of plausible values of $\phi_0$,
produced by quantum fluctuations during inflation, have been
derived by Malquarti and Liddle \cite{phi0}).  In this paper,
we simply emphasize
that many of our results are very sensitive to the assumed value of $\phi_0$.

We now present two representative
examples, the case $V(\phi) = \lambda \phi^2$,
which does not lead to a big rip, and the case $V(\phi)  = \lambda \phi^6$,
which does yield a big rip.
Fig. 1 shows the evolution of the equation
of state parameter $w_{\phi}$ for
$V  = \lambda \phi^2$ with a fixed value of
$\phi_0$ ($\phi_0 = 1$, in the units defined in Sec.
II.A.),
and $\Omega_{m0} = 0.25, 0.30, 0.35$ (where
$\Omega_{m0}$ is the present-day matter density in units
of the critical density).
Note that equation (\ref{Vprime/V}) implies
that for a fixed set of initial conditions,
there is a single functional behavior for $w_\phi$
as a function of $\Omega_\phi$.  Hence, we can simply calculate $w_\phi$
as a function of $\Omega_m$ and define the ``present"
to be the point at which $\Omega_m$ evolves to the given
desired value, $\Omega_{m0}$.  This implies that the three curves for $w(a)$ differ
only in an overall multiplicative constant for $a$; on the
log scale in Fig. 1, the three curves are therefore identical
and are simply displaced horizontally from each other. 
Given this simple dependence on $\Omega_{m0}$, and the fairly
narrow observational limits on $\Omega_{m0}$,
we choose to fix the value of $\Omega_{m0}$
to be $\Omega_{m0} = 0.3$ thoughout the
rest of this paper.

In Fig. 2, we show the evolution of $w_\phi$ for the
same power-law potential for three different values of $\phi_0$ ($\phi_0 = 0.5, 1, 1.5$).
Here we see a much wider divergence in the evolution of these models.
\begin{figure}[t]
\centerline{\epsfxsize=3truein\epsffile{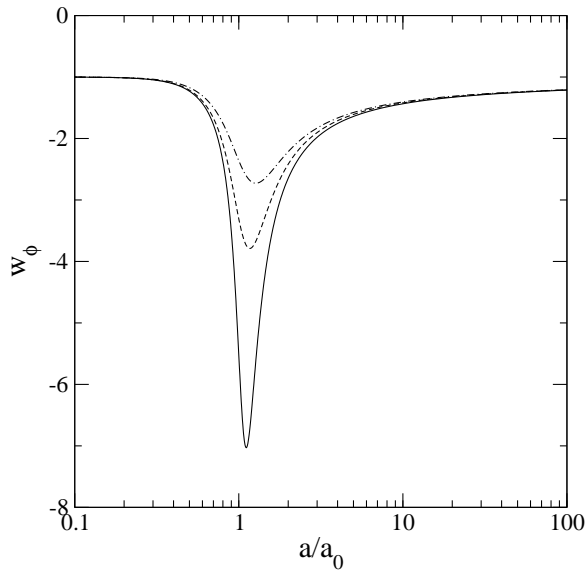}}
\caption{Evolution of the equation of state parameter $w_{\phi}$ as a function of
scale factor $a$ ($a_0$ is the scale factor today)
for $V = \lambda \phi^6$ with
$\Omega_{m0} = 0.3$, and
$\phi_0 = 0.5$ (solid curve), $\phi_0 = 1$ (dashed curve), and $\phi_0 = 1.5$
(dot-dashed curve).}
\end{figure}
Fig. 3 gives the evolution of $w_\phi$
for the power-law potential $V = \lambda \phi^6$
and three different values for $\phi_0$.
We note that for both power laws,
$w_\phi \approx -1$ at early times, but then $w_\phi$
decreases sharply
as soon as $\rho_\phi$ becomes comparable to $\rho_m$.
Then $w_\phi$ asymptotically returns to $-1$ at late times.
This behavior is consistent with the results for the
single power-law model
[$V(\phi) = \lambda \phi^2$] investigated in Ref. \cite{Sami}.

Now consider negative power-law potentials.
In Figs. 4 and 5, we show the evolution of $w_\phi$ for two negative
power-law models:  $V = \lambda \phi^{-1}$ and $V = \lambda \phi^{-2}$.
As for the positive power-law models, these models begin with
$w_\phi \approx -1$, but, as noted in the previous section, $w_\phi$ generically
evolves toward $-\infty$.  The change in behavior from $w_\phi \approx -1$ to
$w_\phi \rightarrow -\infty$ is triggered by $\rho_\phi$
dominating the expansion.  Again, we see that the evolution
is quite sensitive to the initial value of $\phi$.  For small
values of $\phi_0$ ($\phi_0 \lesssim 0.5$), these models
yield a singularity at the present, a result not favored
by observations.
\begin{figure}[t]
\centerline{\epsfxsize=3truein\epsffile{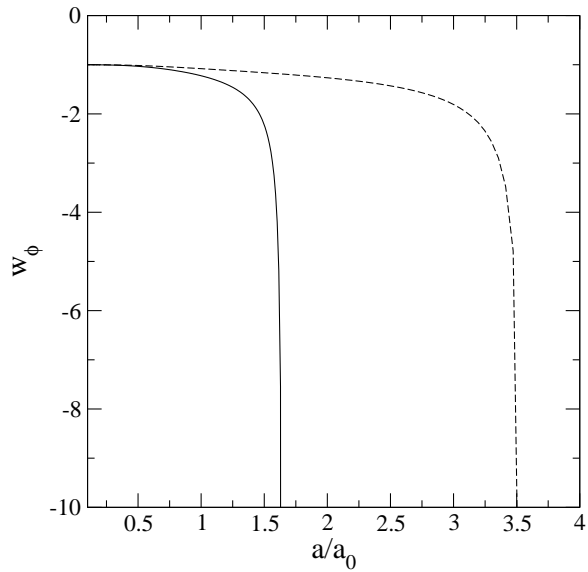}}
\caption{Evolution of the equation of state parameter $w_{\phi}$
as a function of scale factor $a$ ($a_0$ is the scale factor today)
for $V = \lambda \phi^{-1}$
with $\Omega_{m0} = 0.3$ and $\phi_0 = 1$ (solid curve), $\phi =  1.5$ (dashed
curve).}
\end{figure}

\subsubsection{Exponential Potentials}

Now consider a
second class of models of the form
\begin{equation}
\label{Vexp}
V(\phi) = \beta \exp \left( \lambda \phi^{\alpha} \right).
\end{equation}
For simplicity, we will consider only the case
$\lambda > 0$.
As for the power-law case, it is clear from equation (\ref{Vprime/V})
that the functional form of $w_\phi(\Omega_\phi)$
is independent of $\beta$, since $V^\prime/V$ is independent of $\beta$,
so we choose $\beta$ to give
the desired value of $\Omega_{m0}$.

Using our classification scheme of Sec. II.A., we find that
\begin{equation}
\label{Vprimeexp}
\frac{V^\prime}{V} = \lambda \alpha \phi^{\alpha - 1}
\end{equation}
with the field rolling in the positive $\phi$ direction.
Hence, the asymptotic behavior of $w_\phi$ depends on
the value of $\alpha$.
For $\alpha = 1$, equation (\ref{Vexp})
reduces to a simple exponential, previously
investigated in Refs. \cite{ENO,Hao,Sami}.
In this case
$w_\phi \approx -1$ initially, with $w_\phi$ evolving to the
value $w_\phi = -1 - \lambda^2/3$ at late times \cite{Hao}.
As noted in the previous section, this is the only scalar field
model with negative kinetic term that corresponds exactly
to Class 2
($w_\phi = w_0 \ne -1$ at late times).
For $\alpha > 1$,
we see from equation 
(\ref{Vprimeexp})
that $V^\prime/V \rightarrow \infty$ asymptotically,
so that
$w_\phi \rightarrow -\infty$
at late times.
\begin{figure}[t]
\centerline{\epsfxsize=3truein\epsffile{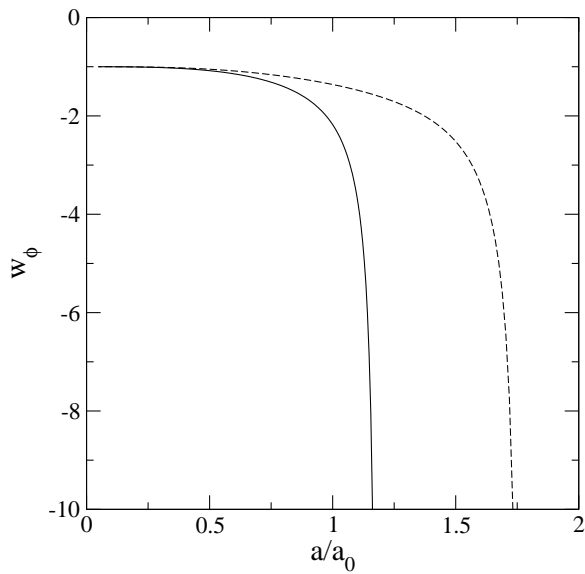}}
\caption{Evolution of the equation of state parameter $w_{\phi}$
as a function of scale factor $a$ ($a_0$ is the scale factor today)
for $V = \lambda \phi^{-2}$
with $\Omega_{m0} = 0.3$ and $\phi_0 = 1$ (solid curve), $\phi =  1.5$ (dashed
curve).}
\end{figure}
(The special case of $\lambda = 1, \alpha = 2$
was investigated in Ref. \cite{Sami}).
Finally, for $\alpha < 1$, we see
that $V^\prime/V \rightarrow 0$ asymptotically, so
that $w_\phi \rightarrow -1$ at late times.  For these models,
the question of whether or not a big rip occurs is
determined by equation (\ref{ripcondition}).  We see that
\be
\int \frac{\sqrt{V(\phi)}}{V^\prime(\phi)} d\phi = 
\int_{\phi_c}^\infty \frac{1}{\lambda \alpha \sqrt{\beta}}
\phi^{1-\alpha} e^{-\lambda \phi^\alpha/2} d\phi
\ee
For $0 < \alpha < 1$, this integral always converges, which
indicates that for these cases we get a big rip.

The behavior of several representative examples of these models is
illustrated in Figs. 6 and 7.
For these models, we take $\phi_0 = 1$, although
the case $\alpha = 1$ (only) is insensitive to the initial
conditions.
For all of these models, the transition in $w_\phi$ is triggered by
the onset of dark energy domination, as was the case for the
power-law potentials.

\section{Supernova Constraints on Phantom Models}

Supernova Ia observations \cite{Knop,Riess} provide the best current constraints
on the value of $w_\phi$.  However, these limits normally assume
either a constant $w_\phi$, or a value for $w_\phi$ that varies linearly with $a$
near the present.  The phantom
models discussed here clearly display quite
different behavior for the evolution of $w_\phi$.  Hence, it is useful to derive
supernova constraints for these models directly.

In this section, we investigate the goodness of fit of various
phantom models to the corresponding observed
luminosity distance $D_{L}^{obs}$ coming
from the SnIa Gold data set \cite{Riess}. 
\begin{figure}[t]
\centerline{\epsfxsize=3truein\epsfbox{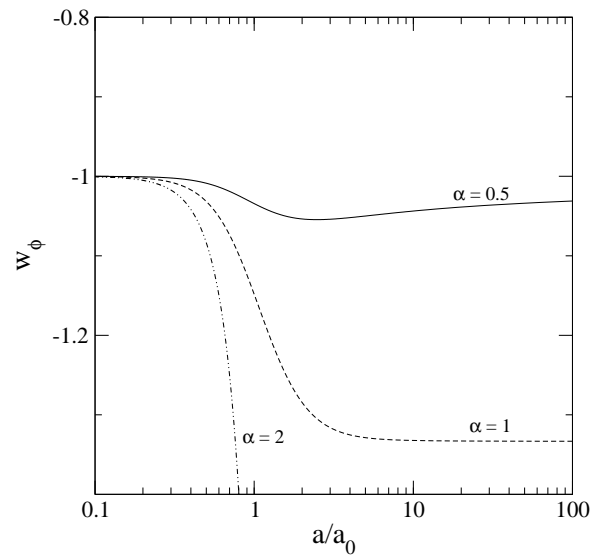}}
\caption{Evolution of the equation of state parameter $w_\phi$ as a function of
scale factor $a$ ($a_0$ is the scale factor today) for the
exponential potential $V = \beta e^{\lambda\phi^\alpha}$, with 
$\lambda = 1$, $\Omega_{m0} = 0.3$, and $\phi_0 = 1$,
for the indicated values of $\alpha$.}
\end{figure}
The observations of supernovae measure essentially the
apparent magnitude $m$, which is related to the luminosity distance $d_L$ by
\begin{align}
m(z) = {\cal M} + 5 \log_{10} D_L(z) ~,
\end{align}
where
\begin{align}
D_L(z) \equiv {\frac{H_0}{c}} d_L(z)~, \label{DL}
\end{align}
is the dimensionless luminosity distance and
\begin{align}
d_L(z)=(1 + z) d_M(z)~,
\label{dL}
\end{align}
with $d_M(z)$ being the comoving distance given by
\begin{align}
d_M(z)=c \int_0^z {\frac{1}{H(z')}} dz'~. \label{dm}
\end{align}
Also,
\begin{align}
{\cal M} = M + 5 \log_{10}
\left({\frac{c/H_0}{1~\mbox{Mpc}}}\right) + 25~,
\end{align}
where $M$ is the absolute magnitude.

The data points in these samples are given in terms of the distance modulus
\begin{align}
\mu_{\rm obs}(z) \equiv m(z) - M(z)~ = 5 \log d_{L} +25,
\end{align}
 where $d_{L}$ is measured in Mpc.
The $\chi^2$ is calculated from
\begin{align}
\chi^2 = \sum_{i=1}^n \left[ {{\mu_{\rm obs}(z_i) - 
\mu_{\rm th}(z_i; H_{0}, c_{\alpha})}\over{\sigma_{\mu_{\rm
obs}}(z_i)}} \right]^2~,
 \label{chisq2}
\end{align}
where the present-day Hubble parameter, $H_{0}$, is a nuisance parameter
that is marginalized over,
and $c_{\alpha}$ are the model parameters.
\begin{figure}[t]
\centerline{\epsfxsize=3truein\epsfbox{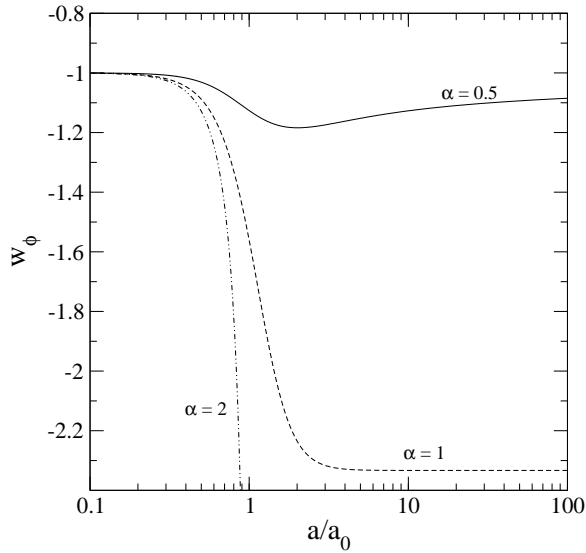}}
\caption{As Fig. 6, for $\lambda = 2$.}
\end{figure}
In what follows, we study $\Delta \chi^{2}$ given by
\begin{equation} 
\Delta\chi^{2} = \chi^{2} - \chi^{2}_{lcdm}
\end{equation}
where $ \chi^{2}_{lcdm} = 177.1$ is the value for the
$\Lambda$CDM model with $\Omega_{m} = 0.3$ and $\Omega_{\Lambda} = 0.7$,
which we take as our fiducial model, as it provides a good fit
to all current cosmological observations. 
Thus, $\Delta \chi^2$ measures how much worse (or better, for
$\Delta \chi^2 <0$) the phantom model fits the data compared to
the standard $\Lambda$CDM model.

The value of $\Delta \chi^2$ for the potentials $V = \lambda \phi^\alpha$
is shown in Fig. 8 as a function of $\alpha$.
(In this section, we take $\phi_0 = 1$ throughout).
It is clear
that large positive or negative values of $\alpha$ are excluded.
At the $2\sigma$ level, we have
\begin{equation}
-1.5 < \alpha < 2.
\end{equation}
A comparison of these limits with the evolution graphs for $w_\phi$
in Sec. II. B. shows, not surprisingly, that the best agreement
with the observations occurs for models which most closely
resemble a cosmological constant up to the present.  The allowed
models with
negative
power-law potentials generically resemble a cosmological constant at early
times, deviating toward large negative values of $w_\phi$ and a singularity
at late times.  In this regard, they resemble some variants
of the Chaplygin gas model \cite{Sen}.  Positive power-law models
with large $\alpha$, which correspond to more negative
values of $w_\phi$ at the present, are excluded.  In particular, we
must have $\alpha > 4$ to produce a big rip, and all such
models are excluded by observations.  Note, however, that
we have confined our investigation to the initial state
$\phi_0 = 1$; these conclusions will differ for other values of
$\phi_0$.

The value of $\Delta \chi^2$ for the exponential
potentials, $V = \beta e^{\lambda \phi^\alpha}$ is shown
in Fig. 9.
The likelihood here is obviously a function
of both $\lambda$ and $\alpha$, but it is clear
that smaller values of $\alpha$ are favored.
Conservative limits are $\alpha < 1.5$
for $\lambda = 1$ and
$\alpha < 1$ for $\lambda < 2$.
Within these limits,
models with $\lambda = 1$
allow all three classes of behavior for the asymptotic value of $w_\phi$,
as defined in Sec. II. A.  However, with $\lambda > 2$, we see
that only $\alpha < 1$ is allowed, corresponding to $w_\phi \rightarrow -1$
asymptotically.  As noted, however, these $\alpha < 1$
models do produce a big
rip at late times.
\begin{figure}[t]
\centerline{\epsfxsize=3truein\epsfbox{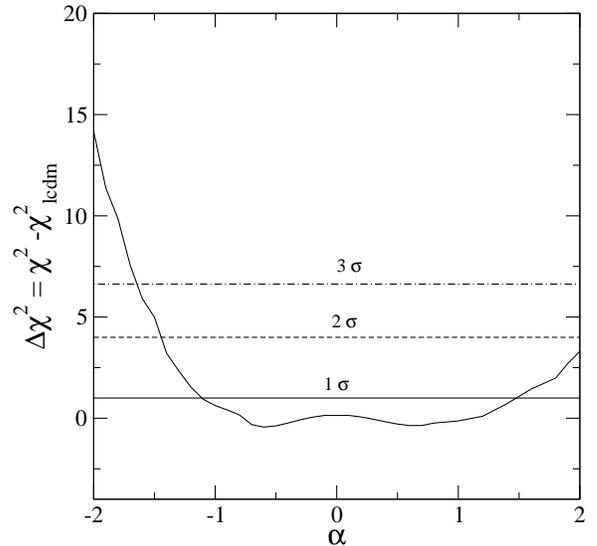}}
\caption{The value of $\Delta\chi^{2} = \chi^{2} - \chi^{2}_{lcdm}$
as a function of $\alpha$
for the power-law potentials $V = \lambda \phi^\alpha$,
with $\Omega_{m0} = 0.3$ and $\phi_0 = 1$.
Here
$\chi^2_{lcdm}$ is the value of $\chi^2$ for the $\Lambda$CDM model
with $\Omega_{m0} = 0.3$ and $\Omega_{\Lambda0} = 0.7$.  Horizontal
lines give the $1\sigma$, $2\sigma$, and $3\sigma$ limits.}
\end{figure}

\section{Coincidence Analysis}

This section is an extension of the coincidence
analysis of Ref. \cite{doomsday} to the models
outlined in the previous section. Observational evidence
suggests that the
matter density and the dark energy density
have similar values today.  However, for a cosmological constant
(for instance), we have $\rho_\Lambda$ constant,
$\rho_m \propto a^{-3}$, so there is only a very small
epoch in time when these two densities are within an order
of magnitude of each other;
this constitutes the ``coincidence problem.''

Scherrer \cite{doomsday} argued that phantom models provide
a natural explanation of this result, since the big rip
in such a universe is itself triggered by the onset of phantom
domination.  Hence, the universe spends an appreciable
fraction of its total lifetime in a ``coincidental" state,
with $\rho_\phi \approx \rho_m$.  (Extensions and elaborations of
this argument can be found in Refs. \cite{Avelino,Cai,Yang}).

As in Ref. \cite{doomsday} we define the parameter $r$ to
be the ratio of scalar field energy density to nonrelativistic
matter density:
\begin{equation}
r \equiv \frac{\rho_{\phi}}{\rho_{m}}. 
\end{equation}
For a universe with a future singularity,
we then calculate the fraction of the total (finite) lifetime of the
universe that is spent in a state for which the two energy densities
are within a factor of $r_0$ from each other:
$1/r_0 < r < r_0$.

This fraction $f$ can be derived by first
calculating the total lifetime of the universe, given by
\be
t_U= \int_0^\infty \frac{da}{a H(a)} \,.
\ee
and comparing it to the
time that the universe spends in expanding from
an initial scale factor, $a_1$, to a final scale factor $a_2$:
\be
t_{12} = t_2-t_1 = \int_{a_1}^{a_2} \frac{da}{a H(a)} \,,
\ee
where we choose the scale factors to correspond to energy density ratios of $r_1 = 1/r_0$ for $a_1$ and $r_2 = r_0$ for $a_2$. 
The coincidence fraction $f$ is then given by
\be
f =  t_{12}/t_U \,,
\ee
which tells us the fraction of time that the energy
densities are within a factor of $r_0$ from each other.
\begin{figure}[t]
\centerline{\epsfxsize=3truein\epsfbox{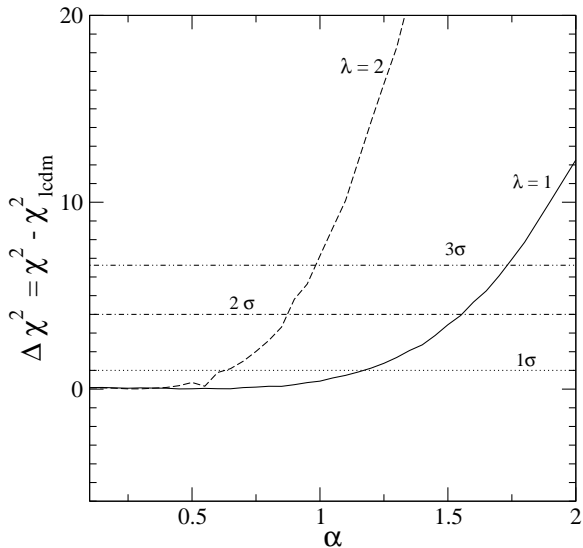}}
\caption{As Fig. 8, for the exponential potentials $V = \beta e^{\lambda
\phi^\alpha}$, with $\lambda = 1$ and $\lambda = 2$.}
\end{figure}
For phantom
models with constant $w_\phi$, this coincidence fraction can be expressed
as \cite{doomsday}
\begin{equation}
\label{fr}
f = \frac{\Gamma(1/2)}{\Gamma(-\frac{1}{2w})\Gamma(\frac{1}{2} +
\frac{1}{2w})}\int_{1/r_0}^{r_0} \frac{r^{-(2w+1)/2w}}
{\sqrt{1+r}} dr.
\end{equation}
A larger value of $f$ corresponds to a universe in which it is
more natural to observe $\Omega_\phi \sim \Omega_m$ today, ameliorating
the coincidence problem.

Note that the choice of $r_0$ is somewhat arbitrary;
in this paper we choose $r_0=10$, corresponding to
$\rho_\phi$ and $\rho_m$ lying within an order of magnitude
of each other.
\begin{figure}[t]
\centerline{\epsfxsize=3truein\epsfbox{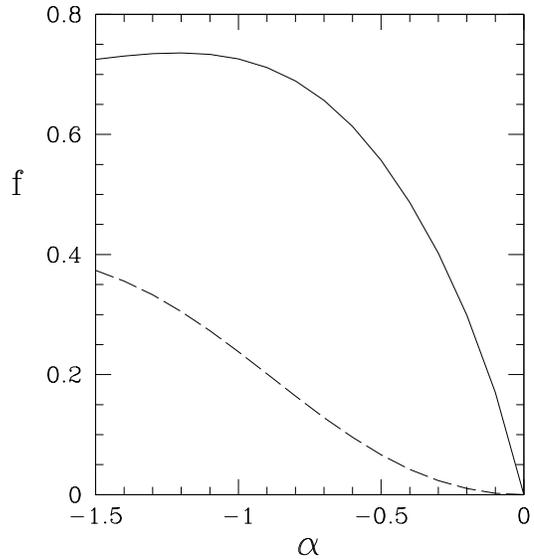}}
\caption{Solid curve gives the fraction of time $f$ that the universe
spends in a state with $0.1 < \rho_\phi / \rho_m < 10$, for phantom
models with $V(\phi) = \lambda \phi^\alpha$; $f$ is given as a function
of $\alpha$.  Dashed curve gives $f$ for models with constant $w_\phi$ having
the same value of $w_\phi$ at the present as the corresponding power-law model.}
\end{figure}
Since our goal is to compare scalar field phantom models with constant-$w_\phi$
phantom models, the actual value of $r_0$ is not crucial.
Further, we note that the value of $f$, like the evolution of
the scalar field, depends on the initial value for $\phi_0$.  For
definiteness, we consider only $\phi_0 = 1$ in what follows, but
these results are easily extended to other values of $\phi_0$.

In Fig. 10, we show the coincidence fraction $f$ for the power-law
models $V(\phi) \propto \phi^\alpha$ as a function of $\alpha$, for
the range of values of $\alpha$ that is not ruled out at the $2\sigma$
level (cf. Fig. 8).  No positive values for $\alpha$ are displayed,
since only the region $\alpha < 2$ is consistent with the supernova
data, while a future singularity can occur
only for $\alpha > 4$, and only models with
a future singularity
are relevant in this coincidence analysis.
Within the allowed range for $\alpha$,
the value of $f$ can be quite large, roughly 0.7 for the most negative
allowed values of $\alpha$.  This means that the universe spends
70\% of its lifetime in a ``coincidental" state, i.e., one in which
the matter and dark energy densities lie within an order of magnitude
of each other.

Further, we note that these scalar-field-based phantom
models provide a more satisfactory solution to the coincidence problem
than the corresponding models with constant $w_\phi$ (such as those
discussed in Ref. \cite{doomsday}) in the sense that the former models
yield a significantly larger value for $f$ for a given current
value of $w_\phi$.  The physical explanation for this lies in the way
in which $w_\phi$ evolves in the models with negative power
law potentials.  In these models, just as in the constant-$w_\phi$ models,
the onset of the singularity is triggered by the dominance of the dark
energy component.  However, in the scalar field models, the beginning
of phantom dominance also triggers a rapid decrease in $w_\phi$, producing an
even more rapid evolution toward the singularity.

In Fig. 11, we show the coincidence fraction $f$ for
the exponential models $V(\phi) = \beta e^{\lambda \phi^{\alpha}}$
for $\lambda = 1$.  Note that this figure encompasses all three
classes of behavior discussed in Sec. II.A.  For $\alpha < 1$, we have
$w_\phi \rightarrow -1$ (with a big rip), for $\alpha = 1$, we have
$w_\phi \rightarrow -4/3$ and for $\alpha > 1$ we have $w_\phi \rightarrow - \infty$.
\begin{figure}[t]
\centerline{\epsfxsize=3truein\epsfbox{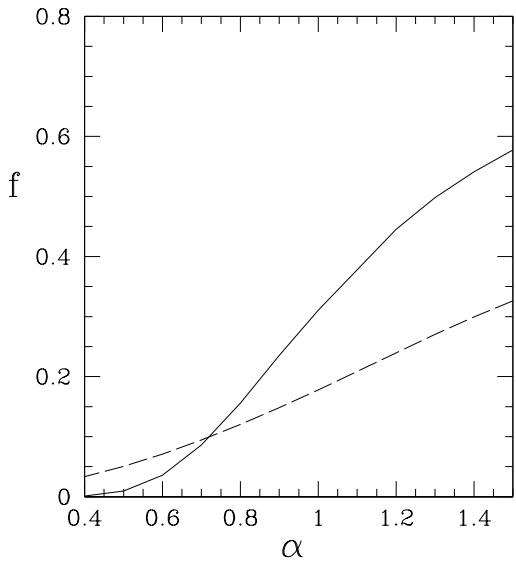}}
\caption{As Fig. 10, for exponential potential $V(\phi) = \beta e^{\lambda
\phi^\alpha}$, with $\lambda = 1$.}
\end{figure}
Again, we see that over most of the allowed range for $\alpha$,
the scalar field models yield a larger value for $f$ than do
the corresponding constant-$w_\phi$ models with the same present-day value
for $w_\phi$.

\section{Conclusions}

Our results provide a comprehensive classification of phantom
models with negative kinetic term for which
$V(\phi) \rightarrow \infty$ asymptotically.  All such models
lead to either
$w_\phi \rightarrow -1$, $w_\phi \rightarrow w_0 < -1$, or $w_\phi \rightarrow -\infty$,
depending on the asymptotic form of the potential.  The first
set of these models need not lead inevitably to a big rip singularity;
we have derived the conditions on $V(\phi)$ which determine when a
big rip occurs.  Our results agree with all of the specific cases that have
been previously investigated elsewhere.

Power-law potentials of the form $V = \lambda \phi^\alpha$ lead to
$w_\phi \rightarrow -1$ for positive power laws and $w_\phi \rightarrow -\infty$
for negative power laws.  However, the supernova observations provide
both upper and lower bounds on $\alpha$.  For the set of initial
conditions on $\phi$ considered here, the only models
which are both
consistent with the observations and
lead to a future singularity
are the negative
power-law potentials.

Generalized exponentials of the form $V = \beta e^{\lambda \phi^\alpha}$
can lead to all three sorts of behaviors indicated above.  We
have $w_\phi \rightarrow -1$ for $\alpha < 1$, $w_\phi \rightarrow w_0$ for
$\alpha =1$ and $w_\phi \rightarrow -\infty$ for $\alpha > 1$.
The supernova observations favor smaller values of $\alpha$;
for the initial conditions examined here, we find that
$\alpha < 1.5$ for $\lambda = 1$ and $\alpha < 1$ for $\lambda = 2$.

These observational limits must be treated with some caution, as
they are highly sensitive to the assumed initial conditions.
For the specific potentials considered here (power laws and generalized
exponentials) we find that the evolution of the scalar field is
very sensitive to the initial value of $\phi$ (but not $\dot \phi$),
in contrast to the ``tracker" quintessence models.  Nonetheless,
these limits do provide a clue to the general forms of the most
observationally-favored models:  not surprisingly, they are the models
which most closely resemble a cosmological constant.

Finally, we have considered the solution of the coincidence problem
in the context of these phantom models.  This argument relies on the
idea that a universe with a future singularity has a finite
lifetime, so it is possible to calculate the fraction of that lifetime, $f$,
for which the densities of the matter and dark energy lie within
an order of magnitude of each other.  Over the observationally-allowed
parameter range, we find that both the power-law potentials and the
exponential potentials yield a larger value for $f$ than do the corresponding
constant-$w_\phi$ models (examined in Ref. \cite{doomsday}) with the same
present-day
value for $w_\phi$.  Thus, these phantom models with time-varying $w_\phi$ provide a better
resolution of the coincidence problem than do constant-$w_\phi$ phantom models.
This is particularly true for models in which $w_\phi$ decreases with time,
as such models lead to a faster evolution toward the singularity.

It is clear that phantom models arising from scalar fields
with negative kinetic terms produce a rich set of behaviors.
The class of such models that is
consistent with current cosmological observations yields
a variety of different possible
future fates for the universe.

\acknowledgements 

R.J.S. was supported in part by the Department of Energy (DE-FG05-85ER40226).

\end{document}